\documentclass[a4paper]{jpconf}

\usepackage{amsmath,amssymb}
\usepackage{graphicx}
\usepackage{xcolor}
\usepackage[colorlinks = true,
linkcolor = magenta,
urlcolor  = blue,
citecolor = red,
anchorcolor = blue]{hyperref}

\begin{document}
	
\title{  Nuclear parton distribution functions (nPDFs) and their uncertainties in the LHC Era  }

\author{ S. Atashbar Tehrani }

\address{ Independent researcher, P.O.Box 1149-8834413, Tehran, Iran }

\ead{atashbart@mail.ipm.ir}
\ead{atashbart@gmail.com}

\begin{abstract}

We have presented the results of our next-to-next-to-leading order (NNLO) QCD analysis of nuclear parton distribution functions (nuclear PDFs) [Phys. Rev. D 93 (2016) 014026, arXiv:1601.00939 [hep-ph]] using all available neutral current charged-lepton ($\ell ^\pm$ + nucleus) deeply inelastic scattering
(DIS) data as well as Drell-Yan (DY) cross-section ratios $\sigma_{\rm DY}^{A}/\sigma_{\rm DY}^{A^\prime}$ for a variety of nuclear targets.
We have studied in detail the parametrizations and the atomic mass (A) dependence of the nuclear PDFs at NNLO at the input scale, $Q_0^2 = 2 \, {\rm GeV^2}$.
Our {\tt KA15} nuclear PDFs provides a complete set of nuclear PDFs, $f_i^{(A, Z)}(x, Q^2)$, with a full functional dependence on $x$, A, $Q^2$.
The uncertainties of the obtained nuclear modification factors for each parton flavour as well as the corresponding observables are estimated using the well-known Hessian method. The nuclear heavy quark contributions are also included into the analysis in the framework of zero-mass variable flavour number scheme (ZM-VFNS).
We compare the results of our parametrization with all available nuclear DIS data and the results of other nuclear PDFs groups. We have found that our nuclear PDFs to be in reasonably good agreement with results in the literature. The estimates of errors provided by our global analysis ({\tt KA15}) are rather smaller than those of other groups.
We have briefly reviewed different aspects of recent LHC heavy-ion collisions data including the first experimental data from the LHC proton+lead ($p-pb$) and lead+lead ($pb-pb$) run which can be used in the global fits of nuclear PDFs.

\end{abstract}
%

\section{ Introduction }

In the past decade, a tremendous experimental efforts have been carried out in order to study the partonic structure of proton.
In spite of the remarkable phenomenological success of Quantum Chromodynamics (QCD) in deep-inelastic-scattering (DIS) experiments, a detailed understanding of the partonic structure of bound nuclei is still lacking.
In the collinear factorized approach to perturbative QCD (pQCD), the structure and dynamics of nucleus are described by the nuclear parton distribution functions (nuclear PDFs)~\cite{Khanpour:2016pph,Eskola:2016oht,Kovarik:2015cma,Wang:2016mzo,Ru:2016wfx,Paukkunen:2017bbm,Aschenauer:2017oxs}.
Analogously to the DIS data available for the free proton case, fixed-target $\ell + A$ DIS scattering has a huge potential to offer information on the nuclear PDFs.
In this work, we have reviewed the results of our recent {\tt KA15} NNLO nuclear PDFs and their uncertainties.
After first introducing to the nuclear DIS data sets, the analysis method are quickly recalled. Then the results of our NNLO nuclear PDFs discussed, and detailed comparison with nuclear DIS experimental data are presented.

\section{ Nuclear DIS data sets }

\begin{figure}[htb]
	\begin{center}
		\vspace{1cm}
		\resizebox{0.45\textwidth}{!}{\includegraphics{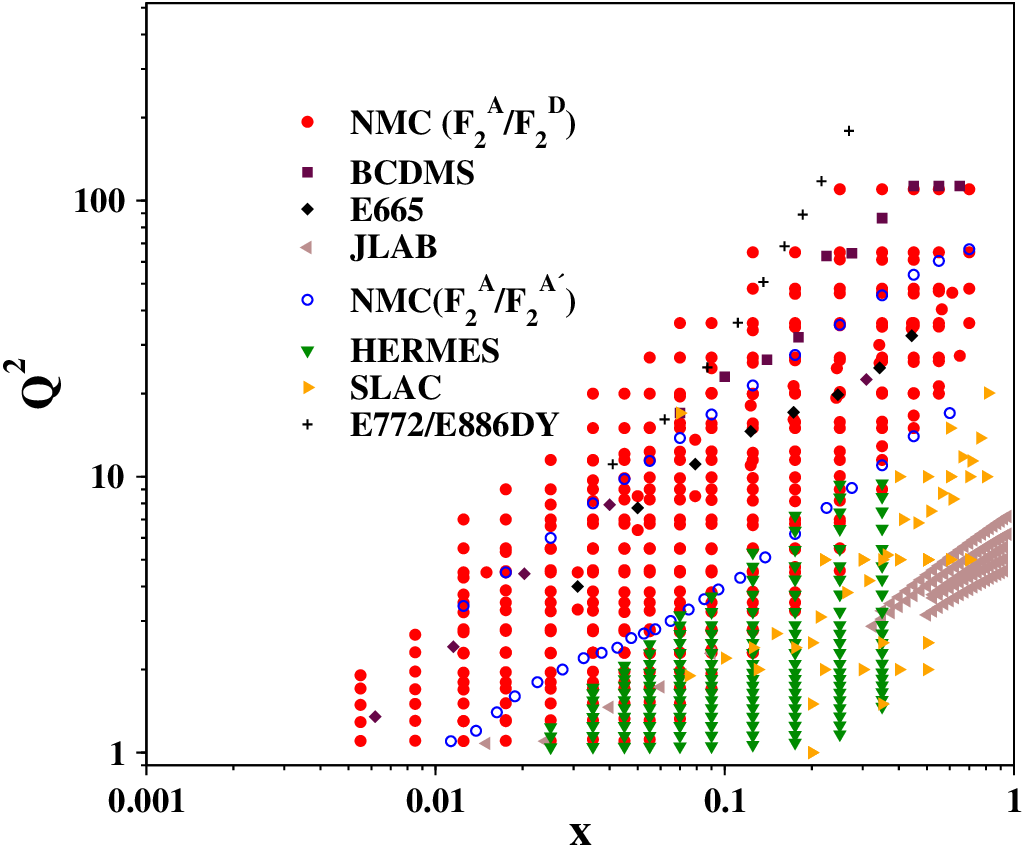}}   
		\caption{ (Color online) Nominal coverage of the data sets used in {\tt KA15} global fits. The plot nicely summarizes the universal $Q^2$ and $x$ dependence of the nuclear DIS data.}\label{fig:xQdata}
	\end{center}
\end{figure}

In our analysis, we have used a large variety of $\ell^\pm-A$ DIS and $pA$ Drell-Yan (DY) data sets. The data sets used in the {\tt KA15} analysis, listed in Ref.~\cite{Khanpour:2016pph}.
The neutral current charged-lepton ($\ell ^\pm$ + nucleus) DIS data as well as DY cross-section ratios $\sigma_{\rm DY}^{A}/\sigma_{\rm DY}^{A^{\prime}}$ used in the {\tt KA15} analysis are shown in Figure.~\ref{fig:xQdata}. This plot nicely summarizes the universal $x$ and $Q^2$ dependence of the nuclear DIS data.
However, the kinematic reach of currently available cross section measurements in $\ell + A$ DIS is much more restricted than in the case of $\ell + p$ DIS.
As a consequence, the nuclear PDFs are significantly less constrained than the proton PDFs.
As one can see from Fig.~\ref{fig:xQdata}, at small value of $x < 0.01$, the obtainable constraints are limited by low statistics.

The LHC $p+pb$ and $pb+pb$ data open a new, high-$Q^2$, kinematic regions for nuclear PDF studies.
Recently, the first global analysis of nuclear PDFs to include LHC proton-lead ($p + Pb$) Run-I data, {\tt EPPS16}~\cite{Eskola:2016oht}, appeared.
They reported the impact of these data on the {\tt EPPS16} nuclear PDFs and shown that the CMS dijets data~\cite{Chatrchyan:2014hqa}, are essential in constraining the nuclear effects in gluon distributions.
The Run-II $p+A$ data as well as Relativistic Heavy-Ion Collider (RHIC)~\cite{Aschenauer:2016our} will have significantly higher luminosities. Therefore, Run-II data are expected to provide much better constraints in the near future.

\section{ Analysis Method }

In {\tt KA15} framework~\cite{Khanpour:2016pph} we parametrize the nuclear PDFs $f_i^{(A,Z)} (x, Q^2)$ which are to be multiplied with the free proton
PDFs $f^p_i(x, Q^2)$, taken here to be those of NNLO JR09 PDFs~\cite{JimenezDelgado:2008hf}.
One can reconstruct the nuclear PDFs as follow at $Q_0^2 = 2 \, {\rm GeV}^2$:
\begin{equation}
f_i^{(A,Z)} (x, Q_{0}^{2}) = w_i(x, A, Z) \,  f^{\rm JR09}_{i}(x, Q_{0}^{2}) \,.
\end{equation}
Here we assume the following functional form for the nuclear modifications
\begin{eqnarray}
w_{i}(x, A, Z) = 1 + \left( 1 - \frac{1}{A^{\alpha}} \right)  \, \frac{a_{i}(A, Z) + b_{i}(A) x + c_{i}(A) x^2 + d_{i}(A) x^3 }{ (1 - x)^{\beta_i} } \,,
\end{eqnarray}
In order to accommodate various nuclear target materials, we introduce
\begin{eqnarray}
b_{i}(A) \rightarrow b_1 A^{b_2};  \, \,  c_{i}(A) \rightarrow c_{1} A^{c_{2}}; \, \, d_{i}(A) \rightarrow d_{1} A^{d_{2}}; \, \, a_{\overline{q}} (A) \rightarrow   a_{1} A^{a_{2}}\,,
\end{eqnarray}
in which \{$p_i = b_1, b_2, c_1, c_2, d_1, d_2, ... $\} are free fit parameters.
To determine the best fit at NNLO, we need to minimize the $\chi^2$ with respect to 16 free input nuclear PDFs parameters in above equations.
The global goodness-of-fit procedure follows the usual chi–squared method~~\cite{Khanpour:2016pph}.  The optimization of the
$\chi^2$ values to determine the best parameters of nuclear PDFs is done by the CERN program library {\tt MINUIT}~\cite{James:1994vla}. The experimental errors are calculated from systematic
and statistical errors, added in quadrature. For the determination of nuclear PDFs uncertainties and corresponding observables, we use the standard Hessian method with a global tolerance $\Delta \chi^2 = 1$. This method are discussed in details in Refs.~\cite{Khanpour:2016pph,Shahri:2016uzl,Martin:2009iq,MoosaviNejad:2016ebo,Khanpour:2016uxh,Harland-Lang:2014zoa,Shoeibi:2017zha,Khanpour:2017fey,Khanpour:2017cha,Shoeibi:2017lrl,Soleymaninia:2017xhc}.
\begin{figure}[htb]
\begin{center}
\vspace{1cm}
\resizebox{0.50\textwidth}{!}{\includegraphics{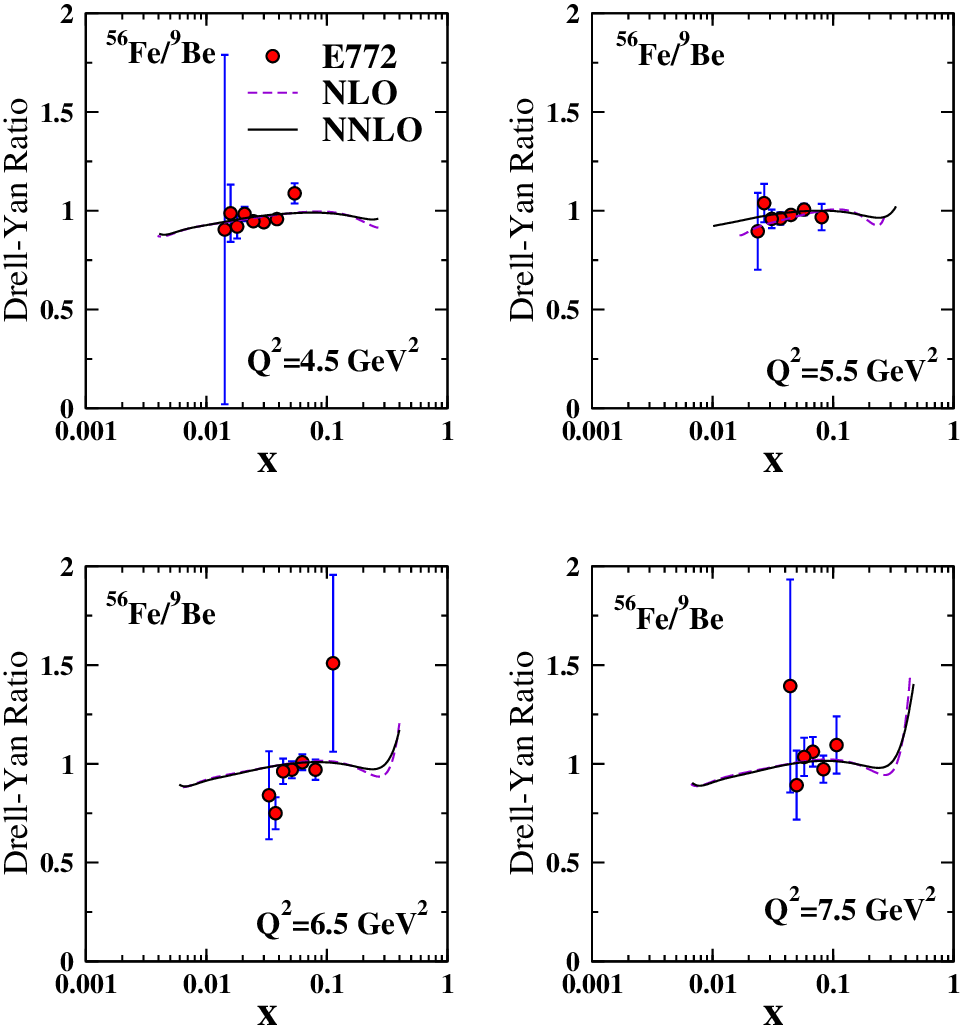}} 
\caption{ (Color online) Our theoretical predictions based on  {\tt AT12} NLO and {\tt KA15} NNLO nuclear PDFs are compared with the data of the DY cross-section ratios $\sigma_{\rm DY}^{Fe}/\sigma_{\rm DY}^{Be}$. Data points are from the FNAL-E866 experiments at Fermilab~\cite{Vasilev:1999fa}.}\label{fig:DYFEtoBe}
\end{center}
\end{figure}

\section{ Results and discussion }

\begin{figure}[htb]
	\begin{center}
		\vspace{1cm}
		\resizebox{0.95\textwidth}{!}{\includegraphics{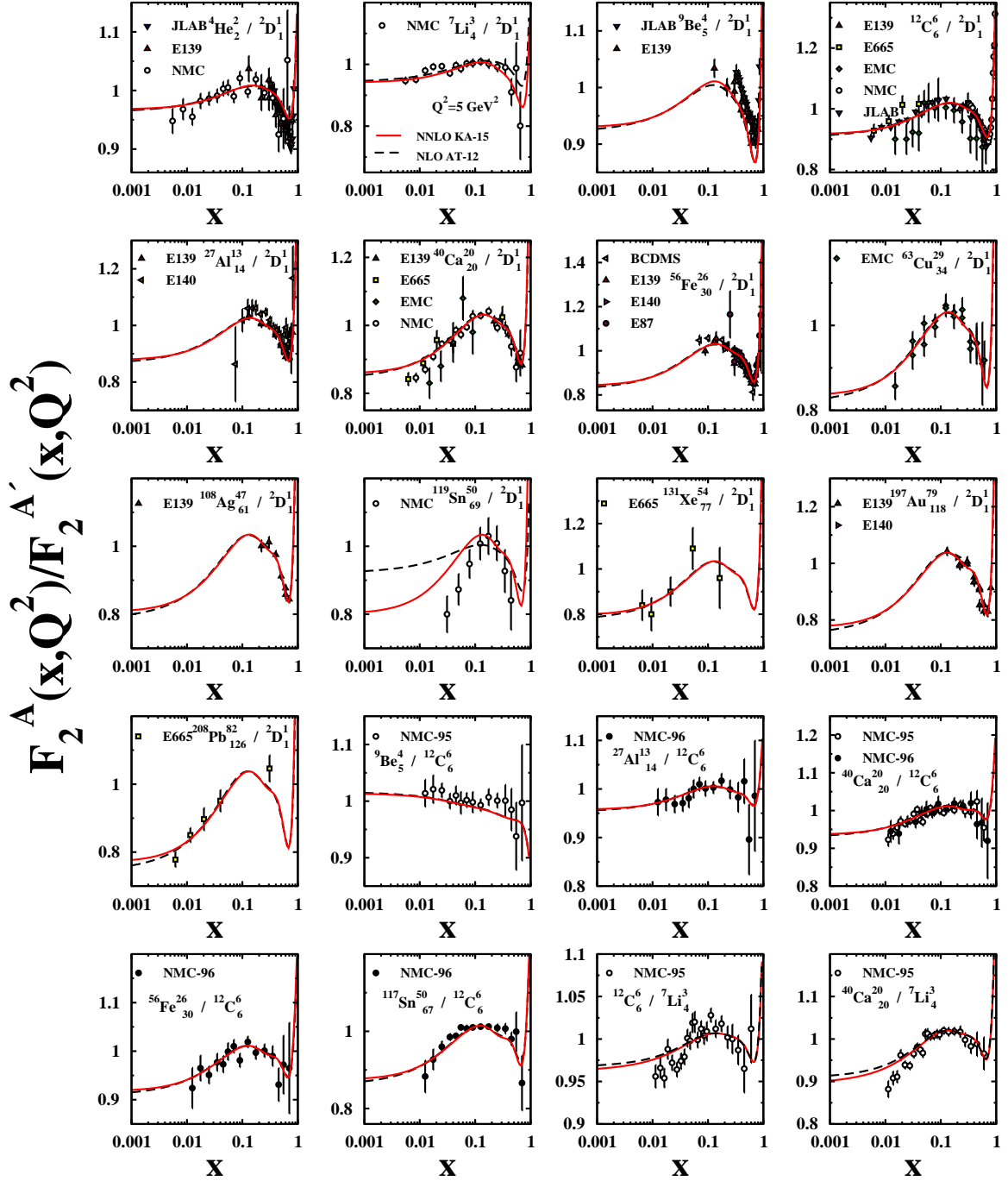}}   
		\caption{ (Color online) Comparison of the {\tt KA15} NNLO~\cite{Khanpour:2016pph} and {\tt AT12} NLO~\cite{AtashbarTehrani:2012xh} theory predictions with the analyzed nuclear DIS data. }\label{fig:Comparison}
	\end{center}
\end{figure}

In this section we present a comparison of the {\tt KA15} fit with the analyzed nuclear DIS data.
A detail comparison of the {\tt KA15} NNLO~\cite{Khanpour:2016pph} and {\tt AT12} NLO~\cite{AtashbarTehrani:2012xh} $x$ dependence theoretical predictions of the structure function ratios $\frac{F_2^A (x, Q^2)}{F_{2}^{A\prime} (x, Q^2)}$ with the analyzed nuclear DIS data are presented in Figure~\ref{fig:Comparison}. In this plot, ratios of structure functions for various nuclei as measured by the JLAB, E139, E140, NMC, and EMC collaborations, compared
with our fit. The error bars shown on the experimental data correspond to the systematic and statistical errors added in quadrature. As one can conclude from the figure, these data set into the {\tt KA15} fit without causing a significant tension.
In Figure~\ref{fig:DYFEtoBe}, our theoretical predictions based on {\tt AT12} NLO and {\tt KA15} NNLO nuclear PDFs are compared with the data of the DY cross-section ratios $\sigma_{DY}^{Fe}/\sigma_{DY}^{Be}$ measured by FNAL-E866~\cite{Vasilev:1999fa}. These plots have shown as a function of $x$ for different Q$^2$ values of 4.5, 5.5, 6.5 and 7.5 GeV$^2$.

As one can see from these plots, the FNAL-E866 data on DY cross-section are in a good agreement with both {\tt AT12} NLO and {\tt KA15} NNLO theory predictions.
As we already mentioned, the data from proton-lead ($p-pb$) and lead-lead ($pb-pb$) collisions at the RUN-I at CERN-LHC would be very desirable in order to determine the nuclear PDFs at low values of parton fractional momenta $x$~\cite{Eskola:2016oht,Salgado:2011wc}.

\section{ Summary and conclusion }

As a short summary, we have presented our global analysis of NNLO nuclear PDFs, {\tt KA15}, extracted from a larger variety
of nuclear DIS data sets. The obtained results are in good agreements with all data analyzed. However, the uncertainties are still significant for all components at small value of $x$ and, clearly, more data is therefore required. In this respect, new data from the $p-pb$ collisions at 13 TeV LHC will be available soon and will provide high-precision DIS constraints for all nuclear parton flavours.

\section*{Acknowledgments}

Author is especially grateful Hamzeh Khanpour for reading this manuscript and for many useful discussions and comments.

\end{document}